# Unsupervised KPIs-Based Clustering of Jobs in HPC Data Centers

**Mohamed S. Halawa** [1,*], **Rebeca P. Díaz Redondo** [2] **and Ana Fernández Vilas** [2]

[1] Business Information Systems Department, Arab Academy for Science Technology and Maritime Transport, Cairo, Egypt;
mhalawa@det.uvigo.es
[2] Information & Computing Lab, AtlanTTIC Research Center, Universidade de Vigo, 36310 Vigo, Spain; rebeca@det.uvigo.es (R.P.D.R.); avilas@det.uvigo.es; (A.F.V.)
* Correspondence: mhalawa@det.uvigo.es;

**Abstract:** Performance analysis is an essential task in High-Performance Computing (HPC) systems and it is applied for different purposes such as anomaly detection, optimal resource allocation, and budget planning. HPC monitoring tasks generate a huge number of Key Performance Indicators (KPIs) to supervise the status of the jobs running in these systems. KPIs give data about CPU usage, memory usage, network (interface) traffic, or other sensors that monitor the hardware. Analyzing this data, it is possible to obtain insightful information about running jobs, such as their characteristics, performance, and failures. The main contribution in this paper is to identify which metric/s (KPIs) is/are the most appropriate to identify/classify different types of jobs according to their behavior in the HPC system. With this aim, we have applied different clustering techniques (partition and hierarchical clustering algorithms) using a real dataset from the Galician Computation Center (CESGA). We have concluded that (i) those metrics (KPIs) related to the Network (interface) traffic monitoring provide the best cohesion and separation to cluster HPC jobs, and (ii) hierarchical clustering algorithms are the most suitable for this task. Our approach was validated using a different real dataset from the same HPC center.

**Keywords:** High performance computing; Time series analysis; Unsupervised learning; Clustering

## 1. Introduction

HPC systems are known for their costly operation and expensive complex infrastructure [1]. Companies and research centers are increasingly demanding this technology to solve different complex computational problems. This has led to a growing need for constant monitoring of HPC systems to ensure stable performance. These monitoring systems are periodically checking the computational nodes of the HPC system to gather the values of different performance counters known as KPIs [2]. This information illustrates the operational status of the system. KPIs are usually organized in different categories, regarding the parameters that are being monitored: CPU usage, Memory usage, network traffic, or other hardware sensors. Each KPI is often recorded as a time series: different values of the same parameter (KPI) that are periodically gathered, with a specific frequency. Thus, KPIs are usually recorded as a time series matrix that can be processed for different purposes: anomaly detection, optimal resource allocation, visualization, segmentation, identifying patterns, trend analysis, forecasting, indexing, clustering, etc. For instance, abnormal behavior in KPIs may explain or predict the existence of some problems like application issues, work overload or system faults in the HPC systems.

Therefore, time series analysis techniques are relevant for the analysis of KPIs. In fact, there are different approaches in the literature [3, 4] based on the analysis of a large number of time varying performance metrics. These proposals apply different techniques, such as statistical analysis [5], machine learning [6, 7], and time series [8]. Among all these approaches, machine learning (ML) stands out in analyzing time series data. The availability of the current advanced ML techniques can quickly process a massive matrix with diverse data types, like text, numerical data, or categorical data. These approaches face some common challenges to analyze the gathered data:

- Large data volume. Each HPC node generates a large number of KPIs (usually more than a thousand). Thus, selecting the most appropriate set of KPIs for job analysis is a key aspect [9].
- Large data dimensionality. The KPI matrix that corresponds to one job may contain a huge number of vectors depending on the number of parallel nodes required during its execution.
- Lack of annotated data. This entails problems to validate the models and methodologies. This problem has been highlighted in previous proposals [10], where only a reduced number of annotated KPIs were used. Consequently, the obtained results cannot be considered complete or representative [10, 11].

Our research work focuses on identifying groups of similar jobs. Since similar jobs tend to have similar performance, we have opted to analyze the KPI data obtained from the monitoring system: each job run in some parallel nodes and the monitoring system is gathering the KPI data per node. We decided to apply clustering techniques to the information given by the KPIs. Besides, the lack of annotated data has driven our research to the application of unsupervised techniques, such as partition and hierarchical clustering algorithms.

As previously mentioned, the large data volume is an important challenge when analyzing the KPIs. So, one of our objectives is identifying which metrics (KPIs) are the most appropriate for clustering. For this to be possible, we have done a two-step analysis. First, we performed clustering by combining KPIs information. Second, we performed clustering using each KPI information individually. The evaluation was done using a real dataset obtained from the Centro de Supercomputación de Galicia (CESGA)[1].

Consequently, our contributions are: (i) a clustering-based methodology that is able to identify groups of jobs that are executed in HPC systems; (ii) simplifying the computational problem by analyzing the different KPIs in order to determine which ones are the most suitable for this type of clustering; and (iii) providing the best clustering algorithm in order to identify different types of HPC jobs according to their performance. This methodology can be applied in any HPC to obtain clusters that identify the different types of running jobs. Finally, the resulting clusters constitute the base for a further analysis that will enable the identification of anomalies in jobs. To the best of our knowledge, this approach entails a novelty approach because of the following aspects: the variety of the KPIs used for our analysis (CPU usage, Memory usage, network traffic and other hardware sensors) and the approach of applying PCA reduction in order to face an overwhelming and challenging clustering of KPIs.

This paper is organized as follows. Section 2 presents some background about the techniques used in this research. Section 3 describes the latest work related to time series clustering and anomalies detection in HPC. Section 4 describes the methodology used in this study. Section 5 defines the experiments and their evaluation. Section 6 provides results discussion and section 7 covers the conclusions and future work proposals.

**2. Background**

There are three types of learning in ML: supervised, semi-supervised, and unsupervised learning. In supervised learning, the data used for analysis is labeled (annotated) before applying any supervised techniques. One example would be a data table with a sequence of behaviors that have labels. This data table is fed to the supervised algorithm to build a model from the labeled data. This model will be used afterward for future predictions. In semi-supervised learning, part of the data is labeled, and the other is not. Finally, in unsupervised learning, the data is not labeled. For example, an unlabeled data table with a sequence of behaviors is fed to an unsupervised algorithm to group the data with similar behaviors with the aim of labeling these groups later [9].

---

[1] CESGA is the supercomputing center in Galicia, a northwest region in Spain (https://www.cesga.es/en/home-2/)

Since we are dealing with a huge number of KPIs that are not labeled, we have decided to consider unsupervised learning techniques and discard other approaches, like classification. In fact, we used clustering techniques that were considered appropriate to discover hidden patterns or similar groups in our dataset without the need of labeled data. In the following subsections, we introduce the algorithms and the distances we have selected (Section 2.1), as well as the different options for clustering validation that helped us to find out the optimal number of clusters (Section 2.2). Finally, we also explained how to deal with a large amount of data by using dimensionality reduction techniques (Section 2.3).

*2.1. Clustering Algorithms*

Clustering algorithms can be classified into five types: partitioning, hierarchical, density-based, grid-based and model-based methods. Since we are interested in applying clustering to a lower dimensional time-series (described in Section 4.3), we have decided to select Partitioning (k-means) and Hierarchical (agglomerative clustering) techniques for clustering as they are the most appropriate for this type of data and widely used for our purpose:

K-means is the most widely used clustering technique thanks to its simplicity. It partitions the data into K-clusters by enhancing the centroids of the clusters and assigning each object in the data to only one cluster. K-means use the Euclidean distance to measure the distance between all the objects and the corresponding centroids to form the cluster [12]. The main advantages of K-means are that it is simple to implement, it is relatively fast in execution, it can be applied in numerous applications that involve a large amount of data, and it obtains very reliable results with largescale datasets [13,14].

Strategies of hierarchical clustering are divided into two types: divisive and agglomerative. Divisive clustering is a "top-down" approach where all objects are initially grouped into one cluster. Then, the objects are split gradually into different clusters until the number of clusters equal to the number of objects. Conversely, the agglomerative clustering is a "bottom-up" approach where each object is assigned to an individual cluster at the initial step of the observation. Then, the clusters are progressively merged until they become one cluster. Agglomerative clustering uses a combination of (i) a linkage method [15, 16] and (ii) a distance metric to merge the clusters. In our analysis, we have used the metrics Euclidean [17], Manhattan [18], and Cosine [19] as well as the following linkage methods:

- Ward's method. It links clusters based on the same function as the K-means (Euclidean distance).
- Single-linkage method. It links clusters based on the minimum distance between two objects of different clusters.
- Complete-linkage method. It links clusters based on the maximum distance between two objects of different clusters.
- Average-linkage method. It links clusters based on the average distance between all the objects of two different clusters.

Hierarchical clustering has important advantages, such as having a logical structure, setting the number of clusters is not required in advance, it provides good result visualization, and it provides dendrogram-based graphical representation [14,20].

*2.2. Cluster Validation*

Many clustering algorithms require the number of desired clusters as an input parameter. Therefore, the experience of the data analyst and/or the specific requirements of the application of the algorithm are keys in determining that number. However, the cluster validation methods are useful to measure the quality of the clustering results and, consequently, to identify the optimal number of clusters. Clustering validation techniques can be classified into two categories: (i) external clustering validation and (ii) internal clustering validation. The former requires -predefined data labels to evaluate the goodness of the cluster, while the latter not require predefined data labels to evaluate the goodness of the cluster [21]. The KPIs of the HPC jobs are usually unlabeled.

Consequently, the internal clustering validation methods are the best option to evaluate the clusters under these circumstances. In fact, our analysis uses three popular internal clustering validation methods to evaluate our clusters: The Silhouette coefficient [22], the Calinski-Harabasz index [21], and the Davies-Bouldin index [23]. These three methods consider for their decision the compactness of the clusters and the separation between them.

The Silhouette index measures the difference between the distance from an object of a cluster to other objects of the same cluster and the distance from the same object to all the objects of the closest cluster. The silhouette score stretches between two values: -1 and 1. The closer the value is to one, the better the shape of the cluster [22]. In fact, a Silhouette score above 0.5 is considered a good result and a result greater than 0.7 is evidence of a very good clustering [24]. Thus, this technique focuses on assessing the shape or silhouettes of the different identified clusters. Besides, the score obtained with this index only depends on the partition, not on the clustering algorithm [22].

$$s(i) = \frac{b(i) - a(i)}{\max\{b(i); a(i)\}} \quad (1)$$

The Calinski-Harabasz index is also identified as a variance ratio criterion, where a cluster validation function is based on the average of the sum of the squared distances among clusters and among objects within the cluster [21]. It focuses on assessing the dispersion of objects within their cluster and the distance from other clusters.

$$CH = \frac{SS_B}{SS_W} \times \frac{(N-K)}{(K-1)} \quad (2)$$

Where $N$ is the total number of samples, $SS_B$ and $SS_B$ are the between and within-cluster variances, respectively, $k$ is the number of clusters.

Finally, the Davies-Bouldin index is used to calculate the separation between the clusters. It focuses on comparing the centroid diameters of the clusters. The closer the Davies-Bouldin value is to zero, the greater the separation is between clusters since zero is the lowest value [23].

$$DB = \frac{1}{K} \sum_{k=1}^{K} \max_{k \neq l} \left\{ \frac{S(u_k) + (u_l)}{d(u_k, u_l)} \right\} \quad (3)$$

Where $S(u_k)+S(u_l)$ is the distance within the cluster and $d(u_k,u_l)$ is the distance between the cluster.

*2.3. Dimensionality Reduction*

HPC KPIs data is usually organized into high-dimensional matrices, which affects the accuracy of any machine-learning algorithms and slows down the model learning process. Hence, it is essential to implement a feature dimension reduction technique that combines the most relevant variables in order to obtain a more manageable dataset [25]. There are several techniques used for dimensionality reduction such as Principal Component Analysis (PCA) [26], t-Distributed Stochastic Neighbor Embedding (t-SNE) [27], and Uniform Manifold Approximation and Projection (UMAP) [28].

The Principal Component Analysis (PCA) [26] is one of the most widely used methods to reduce data dimensionality. Its goal is to reduce data with a large dimension into a small number of the so-called principal components. These principal components highlight the essential features of real data and are expected to maintain the maximum information (variance) of the original data. There are two approaches to apply PCA: (i) fixed PCA and (ii) variable PCA. In the former the number of principal components is fixed beforehand, whereas in the latter the number of principal components is calculated during the process by analyzing the percentage of variance that is maintained.

PCA was successfully applied in different research areas [29, 30, 31, 32, 33]. However, some of them revealed two downsides [25, 27]. On the one hand, in large dimension covariance matrix, the estimation and evaluation tasks are challenging. On the other hand, PCA mainly focuses on the large invariance instead of the small invariance except for the information that is explicitly given in the training data. However, our analysis did not face any of these problems. The maximum dimensionality of the analyzed jobs in our dataset (described in Section 4.2) is 43 parameters. This made the calculation of the principal components feasible with a percentage of retained information greater than 85% for 80% of the jobs (see Section 4.3).

## 3. Related Work

The increasing demand for HPC technology entails that maintaining the quality of the service is key in data centers. Clustering is one of the techniques that is becoming more relevant for this purpose. Analyzing and comparing the differences and similarities of jobs that are run in HPC systems open the door to further and deeper studies, such as anomalies detection. In fact, security and performance go hand by hand. In fact, Zanoon [34] confirmed this direct relationship between security and performance by analyzing the quality of service of cloud computing services (jobs running in HPC systems). The author concluded that better security means better and better performance.

In the specialized literature, there are different approaches that focus on clustering the KPIs in order to support the comparison between jobs [6, 12, 35]. Yahyaoui et al. [12] obtained a good clustering result with a novel approach to cluster performance behaviors. They used different clustering algorithms: K-means, hierarchical clustering, PAM, FANNY, CLARA, and SOM after reducing the dimensionality of time-oriented aggregation of data with the Haar transform.

Li et al. [36] achieved a higher accuracy score for clustering by proposing a robust time series clustering algorithm for KPIs called ROCKA. This algorithm extracts the baseline of the time series and uses it to overcome the high dimensionality problem. Besides, Tuncer et al. [35] proposed a new framework for detecting anomalies in HPC systems by clustering statistical features that retain application characteristics from the time series. On another hand, Mariani et al. [37] proposed a new approach named LOUD that associates machine learning with graph centrality algorithms. LOUD analyzes KPIs metrics collected from the running systems using machine learning lightweight positive training. The objective is twofold: to detect anomalies in KPIs and to reveal causal relationships among them. However, this approach does not work properly with high precision.

## 4. Methodology

HPC systems execute a huge number of jobs every day, which is usually done on hundreds of parallel nodes. These nodes are monitored by more than a thousand KPIs. The goal of this study is to identify clusters of HPC job performances based on the information given by their KPIs. We assume that this task is going to give relevant information about the usual behavior of the jobs, which will be used in the short-term to identify anomalies in jobs. However, this goal brings challenges like data scaling and dimensionality that we have faced defining a six-step methodology which is summarized in Figure 1.

The first step was the selection and definition of the KPIs used in clustering (Section 4.1). The second step was data preprocessing (Section 4.2), where we managed to read the data and identify the jobs that were used in operational jobs, which are those that have a systematic nature like scheduled system update, sensors checks, and backups. On the other hand, non-operational jobs are those that have a non-systematic nature. In addition, a basic analysis of non-operational jobs gave us a better view of the data to prepare them for the pre-clustering phase.

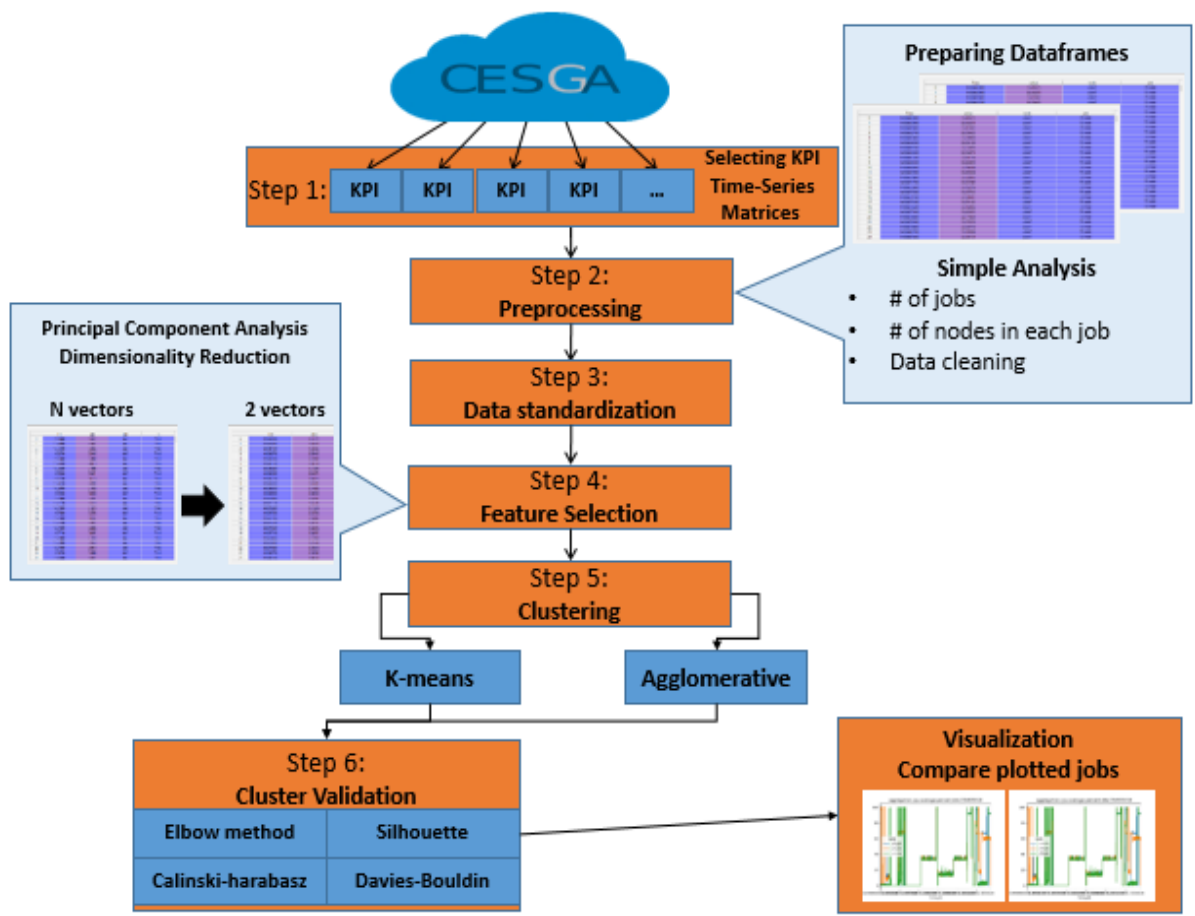

**Figure 1.** Framework for clustering HPC jobs KPIs using feature selection

Some of the dimensionality reduction methods applied like PCA were affected by the scale, which is a requirement for the optimal performance of many machine-learning algorithms. For this reason, a third step to standardize data was needed (Section 4.2). The fourth step was to overcome the dimensionality problem (Section 4.3), always present when analyzing large time series data, like in our case. The PCA dimensionality reduction method helped to reduce our KPIs matrix and speed up the clustering process. The fifth step was clustering (Section 4.4). Two clustering experiments were performed using K-means and agglomerative hierarchical algorithms with different linkage methods and distance metrics (Section 5). The first experiment clustered the PCAs of the non-operational jobs for all the metrics (KPIs) combined. The second experiment clustered the PCAs of the non-operational jobs for each KPI individually. The study did not have a predetermined number of clusters (K). Therefore, in the sixth step, both algorithms clustered the data considering different values of K (from 2 to 200). Then, the clustered results of all K values were evaluated using three previously mentioned internal cluster validation methods (Silhouette analysis, the Calinski-Harabasz index, and the Davies-Bouldin index) to determine the goodness of the clusters and to identify the optimal number of clusters. The clustering results from both experiments were compared to identify which KPIs show the best clustering results and, consequently, are the most representative to cluster the jobs. Lastly, a validation experiment was conducted with a new dataset to validate the obtained results.

*4.1. Performance Data Selection*

The execution of HPC jobs is deployed over a high number of nodes, thousands of parallel nodes that are closely monitored by specific systems. As previously mentioned, these monitoring systems are periodically gathering the values of specific metrics or KPIs. Depending on the monitoring system, the information may be overwhelming with thousands of metrics or KPIs. The collected data is stored as time series matrix per node. These KPIs are usually classified into five different categories:

- Metrics about CPU usage, such as the time spent by a job in the system, owner of the job, nice (priority) or idle time.
- Metrics of network (interface) traffic, such as the number of octets sent and received, packets and errors for each interface.
- IPMI (Intelligent Platform Management Interface) metrics that collect the readings of hardware sensors from the servers in the data center.
- Metrics about the system load, such as the system load average over the last 1, 5 and 15 minutes.
- Metrics of memory usage, such as memory occupied by the running processes, page cache, buffer cache and idle memory.

For our analysis, we have acquired a dataset from the CESGA Supercomputing Center (Centro de Supercomputación de Galicia). Foundation CESGA is a non-profit organization that has the mission to contribute to the advancement of Science and Technical Knowledge, by means of research and application of high performance computing and communications, as well as other information technologies resources. The dataset stores information about a total amount of 1,783 jobs (operational jobs and non-operational jobs), which were running in the 74 available parallel nodes from 1st June 2018 to 31st July 2018.

The collected data give information about 44,280 different KPIs. In order to filter this overwhelming amount of data, we have done a previous filter according to the needs of the CESGA experts. Therefore, we focus our attention on the 11 KPIs summarized in Table 1. The selected KPIs belong to the five previously mentioned categories (CPU usage, memory usage, system load, IPMI and network interface), and were selected by the CESGA experts based on their relevance and clear representation of the performance of jobs from each category.

Each KPI gives a matrix with the following information: (i) the value of the KPI, (ii) the time of the machine when the value was acquired, (iii) the job and (iv) the node to which this value belongs.

**Table 1.** Performance metrics selected

| Category | Metric | Definition |
| --- | --- | --- |
| CPU usage | aggregation.cpu-average.percent.idle | The aggregated average Percent of time when CPU is idle. |
| | aggregation.cpu-average.percent.system | The aggregated average Percent of time when CPU is working. |
| | aggregation.cpu-average.percent.wait | The aggregated average Percent of time when CPU is waiting |
| Network (interface) traffic | interface.bond0.if_octets.rx | The number of bytes received over the network per second. |
| | interface.bond0.if_octets.tx | The number of bytes transmitted over the network per second. |
| IPMI | ipmi.CPU1_Temp ipmi.CPU2_Temp ipmi.PW_consumption ipmi.System_Temp | The temperature readings of CPU1 The temperature readings of CPU2 The power consumed by the system hardware The temperature readings of system |
| System Load | load.load.shortterm | System load average over the last minute |
| Memory usage | memory.cached.memory | Cached Memory occupied |

*4.2. Data Preprocessing and Standardization*

The objective of this preprocessing phase was to read and organize the KPI matrices into data frames before applying any machine-learning steps. For this task, we used the functionality of the Python Pandas library [38]. Additionally, we have also done analysis and data visualization that helped understand the nature of our dataset before applying any further analysis, whose results are summarized in Table 2.

From a total of 1,783 jobs, 200 were excluded from our clustering analysis because of one of the following reasons:
- The jobs were not included in all the 11 KPIs matrices, i.e. we do not have complete information about the metrics of the job.
- The jobs were executed in only one node, which entails they were not parallelized jobs, which is mandatory for our proposed method dimensionality reduction phase.
- These one-node jobs (12% of the dataset) were mostly operational jobs, which are not the focus of our study.
- The analysis of one-node jobs (operational) deserves a specific study that is out of the scope of this paper.

Before proceeding to job clustering, we split the 1,583 jobs into two types: operational -totaling 1,281 jobs- and non-operational –totaling 302 jobs. As it was previously mentioned, our analysis focused only on non-operational jobs. Consequently, we ran two clustering experiments considering the 302 non-operational jobs. In the first experiment, clustering the 11 KPI matrices combined and, in the second experiment, clustering each KPI matrix individually.

**Table 2.** Basic data analysis

| | | |
|---|---|---|
| **Total number of jobs** | | 1,783 |
| **Jobs** | **Operational** | 1,281 |
| | **Non-operational** | 302 |
| **Jobs excluded because they contain less than two nodes** | | 144 |
| **Jobs excluded because they were not included in all the 11 KPI matrices** | | 56 |

Table 3 shows the number of nodes per non-operational job in our dataset. The executable nodes count per job revealed the following: zero jobs were executed on only one node, 195 jobs were executed on less than 5 nodes and 49 jobs were executed on nodes in between 6 and 10. Finally, the calculation showed that 80.7% of the jobs were executed on less than 10 nodes.

**Table 3.** Number of nodes per job (non-operational)

| Node ranges | Number of Jobs |
|---|---|
| 2-5 | 195 |
| 6-10 | 49 |
| 11-15 | 15 |
| 16-20 | 13 |
| >20 | 30 |
| Total | 302 |

The standardization process is usually a required step before applying any machine learning algorithm, in order to achieve reliable results [39]. In our case, we proceeded to do this standardization stage because PCA is affected by scale and the values gathered in the 11 KPI matrices ranged from very low to very high values. Thus, the data was standardized into a unit scale: the mean is equal to zero and the variance is equal to one.

*4.3. Jobs KPIs: Dimensionality Reduction*

One of the major challenges in KPIs analysis is the large volume of available data. After pre-processing our dataset, each column of the matrix represents the KPIs of the nodes that are being used to run the jobs in parallel. The number of nodes is proportional to the parallelization and computational needs of each job as (Time x Nodes) matrix. Analyzing our data, we can see that 19.3% of the jobs were executed on more than 10 nodes. We also have the time series storing the KPIs for each node, so the analysis of such volume of data is overwhelming. Consequently, we have decided to apply a dimensionality reduction method to overcome this challenge. As previously mentioned, we decided to use PCA to reduce the dimensionality of the matrix that represents the KPI gathered data of each job. The objective is reducing this dimensionality without losing information (variance) and, therefore, reducing the computation load and execution time of the clustering algorithms.

We decided to apply a fixed PCA technique with two principal components. This decision is based on two aspects. On the one hand, our initial analysis (Section 4.2) showed that 195 jobs of the total have from two to five nodes. Moreover, 80.7% of the jobs were executed on less than 10 nodes. Thus, applying more than two principal components does not seem to be appropriate in this context. On the other hand, we have checked that applying two principal components was enough to retain information (variance) of the original data (job KPIs performance): the percentage of retained information is greater than 85% in 81% of the jobs, as Table 4 shows.

The PCA was applied to each KPI matrix individually resulting in a matrix of (time x 2 principal components) for each job. On the one hand, for experiment one (Section 5.1), we used jointly the information of the 11 KPIs. For this, we took advantage of the Python Pandas library [38] to combine and flatten the PCA results of each job for the all-11 KPIs into one row in a data frame labeled with the job number resulting in a matrix of (jobs x (times x 2 principal components x KPIs)). Each row in this data frame represents the PCAs for all 11 metrics combined with each job indexed by job number. On another hand, for experiment two (Section 5.2), we analyze each KPI individually. Thus, the PCA results of each job for each KPI were combined and flatten into one row in a separate data frame labeled with the job number resulting in a matrix of (jobs x (times x 2 principal components).

**Table 4.** PCA two principal components retained information

| Retained information ranges percentages | Number of Jobs |
|---|---|
| >= 95% | 142 |
| >=90% - <95% | 76 |
| >=85% - <90% | 27 |
| >=80% - <85% | 26 |
| >=75% - <80% | 20 |
| <75% | 11 |

*4.4. Clustering*

The study applied the K-mean algorithm and the agglomerative hierarchical algorithm to cluster the jobs for both experiments. On the one hand, the K-means used only Euclidean distance for clustering. On the other hand, the agglomerative hierarchical algorithms used three distance metrics -Euclidean, Manhattan, and Cosine- with different linkage methods for clustering. Both algorithms were applied with different numbers of iterations for the number of clusters -from 2 to 200- because no predetermined number of clusters (K) was given. All clustering results were stored and evaluated with three internal cluster validation methods: the silhouette score, the Calinski-Harabasz index and the Davies-Bouldin index, to determine the optimal number of K for the K-means and the agglomerative hierarchical algorithms using all distances. Figure 2(a) and Figure 2(b) illustrate the scores of each cluster for each clustering validation methods, Silhouette score (a) and Davies-Bouldin index (b), to identify the optimal number of clustering visually. In Figure 2(a), a Silhouette score close to 1 implies a better cluster shape. On the contrary, in Figure 2(b), a Davies-Bouldin index close to zero implies greater separation between clusters as described in Section 2.2.

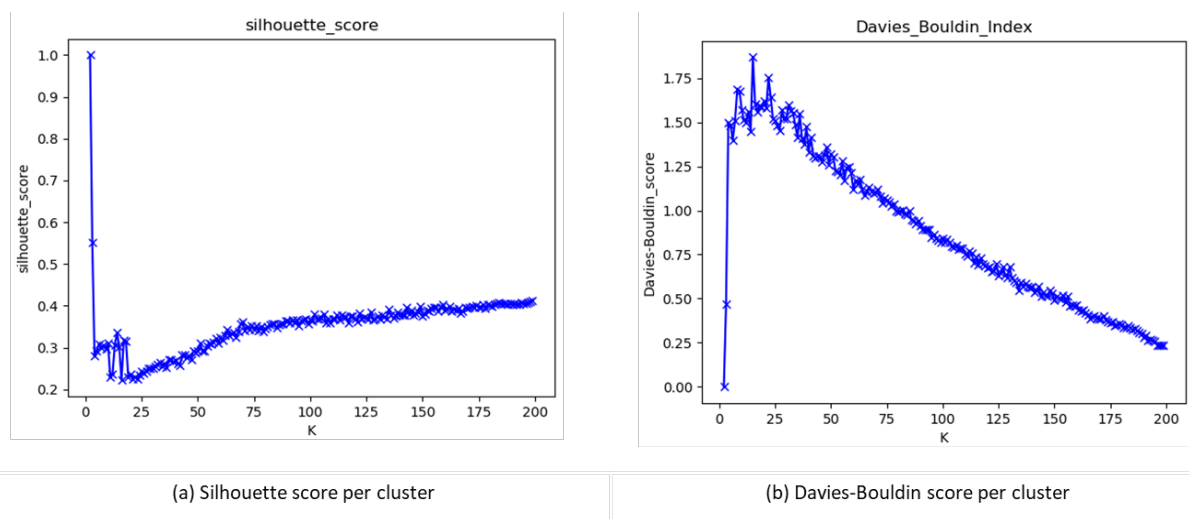

| (a) Silhouette score per cluster | (b) Davies-Bouldin score per cluster |

**Figure 2.** Clustering quality scores

## 5. Experiment Results

*5.1. Experiment One: Results*

In this experiment, we clustered all the non-operational jobs taking into account the information provided by the 11 KPIs. With this aim, we have applied the k-means algorithm and the agglomerative hierarchical algorithm with different linkage rules, as shown in the experimental set-up in Table 5. We did not have a predetermined number of clusters for both algorithms. The clustering was done with a number of iterations for K from 2 to 200 and the results were fed to the three cluster validation methods to identify the optimal number of clusters.

**Table 5.** Experiment one set-up

| Number of jobs | Clustering Algorithm | Cluster Validation Methods |
|---|---|---|
| 302 Non-operational jobs | K-means, Agglomerative Hierarchical | Silhouette score<br>Calinski-Harabasz index<br>Davies-Bouldin index |

Table 6 illustrates the comparison of the optimal numbers of clustering for both algorithms using each one of the three validation methods. Regarding the combined selected 11 KPIs jobs values, we found that the agglomerative hierarchical algorithm performance is better than the K-means algorithm using the Euclidean distance average linkage with a Calinski-Harabasz score of 24,545,720,615 and a silhouette score of 0.523 for 3 clusters. The combined selected 11 KPIs jobs values also performed well with the hierarchical single-linkage clustering using the Euclidean distance, with a Davies-Bouldin score of 0.503 for 13 clusters.

Table 6. Experiment one: results

|  |  |  |  | Calinski-Harabasz Index | Davie-Bouldin Index | Silhouette Score |
|---|---|---|---|---|---|---|
| K-mean | | Euclidean | K | 8 | 6 | 6 |
| | | | Score | 14591690919 | 1.399 | 0.313 |
| Agglomerative Hierarchical | Cosine | Average | K | 8 | 11 | 8 |
| | | | Score | 11439074172 | 1.721 | 0.236 |
| | | Complete | K | 14 | 12 | 11 |
| | | | Score | 6395676953 | 1.916 | 0.176 |
| | | Single | K | 3 | 8 | 3 |
| | | | Score | 22147812202 | 1.173 | 0.074 |
| | Euclidean | Ward | K | 4 | 4 | 4 |
| | | | Score | 22147812202 | 0.961 | 0.069 |
| | | Average | K | **3** | 9 | **3** |
| | | | Score | **24545720615** | 0.607 | **0.523** |
| | | Complete | K | 8 | 24 | 8 |
| | | | Score | 14665419924 | 1.414 | 0.288 |
| | | Single | K | 12 | **13** | 13 |
| | | | Score | 6787242293 | **0.503** | 0.368 |
| | Manhattan | average | K | 13 | 12 | 13 |
| | | | Score | 8253129705 | 0.726 | 0.384 |
| | | complete | K | 4 | 4 | 4 |
| | | | Score | 19996681144 | 1.193 | 0.314 |
| | | single | K | 16 | 9 | 16 |
| | | | Score | 15670931203 | 0.5933 | 0.270 |

*5.2. Experiment Two: Results*

In this experiment, we clustered all the non-operational jobs using only one of the KPIs each time. That is, the study had performed 11 clustering procedures. Once one of the KPIs is selected, the procedure is the same as in experiment one: using the k-means algorithm and the agglomerative hierarchical algorithm with different linkage rules -see the experiment set-up in Table 7. Without a predetermined number of clusters for both algorithms, the number of iterations considered for K ranged from 2 to 200 as done in the previous experiment. Then the results were fed to the cluster validation methods to identify the optimal number of clusters.

Table 7. Experiment two set-up

| KPI | 302 operational jobs considering each KPI individually | Clustering algorithm | Cluster Validation Methods (VM) |
|---|---|---|---|
| 1 | aggregation.cpu-average.percent.idle | | Calinski-Harabasz Index (C) |
| 2 | aggregation.cpu-average.percent.system | | |
| 3 | aggregation.cpu-average.percent.wait | | |
| 4 | interface.bond0.if_octets.rx | K-means, Agglomerative hierarchical | Davies Bouldin Index (D) |
| 5 | interface.bond0.if_octets.tx | | |
| 6 | ipmi.CPU1_Temp | | |
| 7 | ipmi.CPU2_Temp | | |
| 8 | ipmi.PW_consumption | | Silhouette score (S) |
| 9 | ipmi.System_Temp | | |
| 10 | load.load.shortterm | | |
| 11 | memory.cached.memory | | |

The results of clustering each of the 11 KPIs individually showed that the K-means performed well using the Euclidean distance. The results give a Calinski-Harabasz score of 726.341 for 4 clusters in the KPI interface.bond0.if_octets.tx, as shown in Figure 3.

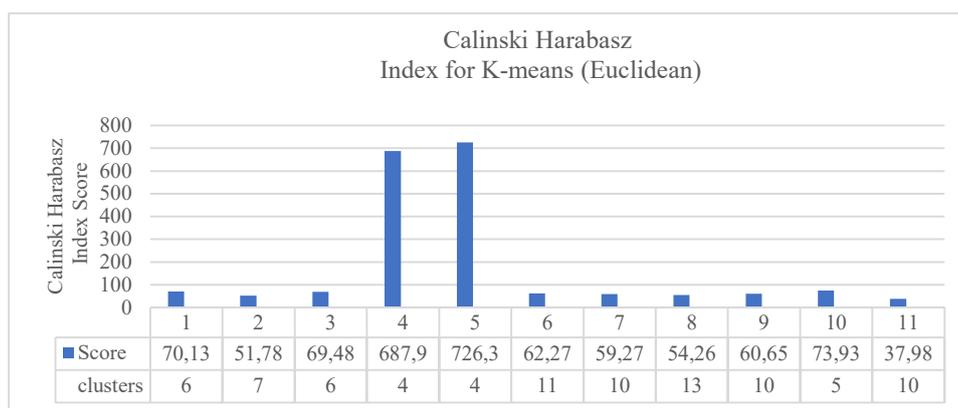

Figure 3. The results of Calinski-Harabasz scores for k-means Euclidean distance.

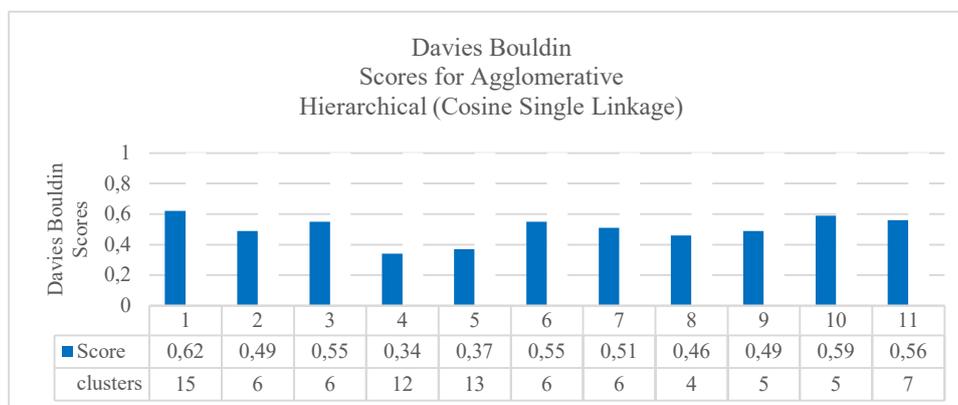

Figure 4. The results of Davies-Bouldin scores for Agglomerative Hierarchical Cosine Single Linkage.

Additionally, the results confirm that the agglomerative hierarchical algorithm performed well in clustering jobs. Figure 4 shows the results with cosine distance; single linkage and Davies-Boulding index. The results show a good score (0.340) using the KPI interface.bond0.if_octets.rx with 12 clusters. Figure 5 shows the results with Manhattan distance; average linkage and Silhouette index. The results show a good score (0.598) using the KPI interface.bond0.if_octets.rx with 4 clusters. All the results are summarized in a complete Table (Table A1) in Appendix A.

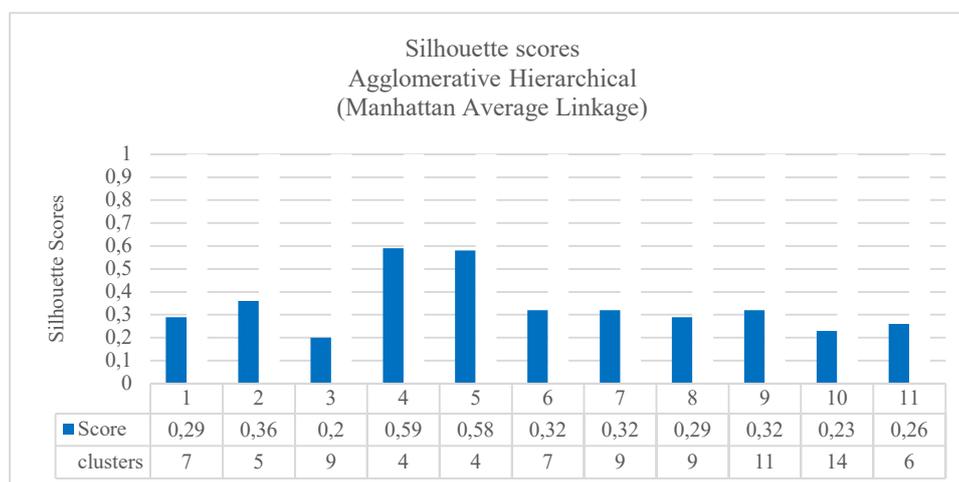

**Figure 5.** The results of Silhouette scores for Agglomerative Hierarchical Manhattan Average Linkage.

*5.3. Validation Experiment*

With the aim of validating the conclusions obtained --KPIs belonging to the Network interface traffic are the most adequate to obtain a good clustering of non-operational jobs that run in the HPC system--, we have performed a new experiment with a different dataset also acquired from CESGA. We have used the same methodology used in experiments one and two (data preprocessing, data standardization, dimensionality reduction, and clustering), but using only the information about the two selected KPIs: interface.bond0.if_octets.rx, and interface.bond0.if_octets.tx.

The dataset stores information about a total amount of 1,500 jobs (non-operational jobs), which were running in the 81 available parallel nodes from 1st August 2019 to 31st September 2019. Table 8 shows the number of nodes per job (non-operational) in the new dataset.

**Table 8.** Number of nodes per job (non-operational) in validation experiment new dataset

| Node ranges | Number of Jobs |
|---|---|
| 2-5 | 856 |
| 6-10 | 320 |
| 11-15 | 74 |
| 16-20 | 56 |
| >20 | 194 |
| Total | 1500 |

The results of clustering based on these two KPIs are shown in Table 9. The highlighted scores in Table 10 demonstrates the best results of the comparison between the scores of the three clustering validation methods for all clustering algorithm. This implies that interface.bond0.if_octets.tx KPI showed better clustering results in all measures cluster shape, cohesion, and separation than interface.bond0.if_octets.rx KPI in the performance of both algorithms (K-means and agglomerative hierarchical) with different distance metrics and linkage methods. K-means performed well using the Euclidean distance with Calinski-Harabasz score of 4,608.5 for 3 clusters; the agglomerative hierarchical algorithm performed well in clustering jobs with cosine distance; single linkage of Davies-Boulding score 0.119 for 3 clusters and Manhattan distance; complete linkage with Silhouette score of (0.858) with 3 clusters using the KPI interface.bond0.if_octets.rx.

Table 9. Validation experiment: results

| | | | | interface.bond0.if_octets.rx | | | interface.bond0.if_octets.tx | | |
|---|---|---|---|---|---|---|---|---|---|
| | | | | Calinski-Harabasz Index | Davie-Bouldin Index | Silhouette Score | Calinski-Harabasz Index | Davie-Bouldin Index | Silhouette Score |
| K-mean | Euclidean | | K | 2 | 2 | 2 | 3 | 2 | 5 |
| | | | Score | 2935.4 | 0.6332 | 0.6545 | **4608.5** | 0.331 | 0.779 |
| Agglomerative Hierarchical | Cosine | Average | K | 4 | 2 | 4 | 3 | 3 | 3 |
| | | | Score | 1893.5 | 0.3284 | 0.6636 | 2321.4 | 0.138 | 0.8548 |
| | | Complete | K | 5 | 2 | 5 | 4 | 2 | 3 |
| | | | Score | 1764.8 | 0.3284 | 0.6508 | 2268.6 | 0.149 | 0.8562 |
| | | Single | K | 5 | 2 | 17 | 4 | 3 | 3 |
| | | | Score | 747.9 | 0.3284 | 0.6242 | 2336.4 | **0.119** | 0.8508 |
| | Euclidean | Ward | K | 2 | 3 | 3 | 2 | 2 | 6 |
| | | | Score | 2882.2 | 0.6399 | 0.6627 | 4569.9 | 0.3589 | 0.7810 |
| | | Average | K | **4** | 2 | **3** | **12** | **2** | **3** |
| | | | Score | 2010.0 | 0.3284 | 0.6532 | 3116.9 | 0.1492 | 0.8563 |
| | | Complete | K | 11 | 2 | 3 | 2 | 2 | 3 |
| | | | Score | 2603.9 | 0.3284 | 0.6502 | 2351.0 | 0.1492 | 0.85841 |
| | | Single | K | 7 | **2** | 8 | 4 | 3 | 3 |
| | | | Score | 1274.4 | 0.3284 | 0.6176 | 2336.4 | 0.1199 | 0.85088 |
| | Manhattan | average | K | 5 | 2 | 3 | 12 | 2 | 3 |
| | | | Score | 2010.0 | 0.3284 | 0.6238 | 3122.1 | 0.1492 | 0.85634 |
| | | complete | K | 8 | 2 | 3 | 2 | 2 | 3 |
| | | | Score | 2143.8 | 0.3284 | 0.6365 | 2351.0 | 0.1492 | **0.85848** |
| | | single | K | 8 | 2 | 8 | 4 | 3 | 3 |
| | | | Score | 1221.9 | 0.3284 | 0.61760 | 2336.4 | 0.1199 | 0.8508 |

**6. Discussion**

After obtaining the results from both experiments shown in Table 6 and Table A1, we have done two comparisons. The first one is done between the results of experiment two to identify which KPI provides the best clustering results in terms of cohesion and separation. With this aim, we have analyzed the results obtained from all the experiments that have been done taking into account the information given individually per KPI (different clustering methods, different metrics, different linkage methods and the assessment with the three quality indexes). The second one is done between the results of experiment one and experiment two to identify which is the best clustering approach, according to the quality indexes. With this aim, we have compared the clustering results when we take into account the joint information given by the 11 KPIs together and the results obtained with the KPI that offered the best result in the first comparison.

The results of the first comparison showed that the results obtained by using the KPI interface.bond0.if_octets.rx and the KPI interface.bond0.if_octets.tx are the best ones with different quality indexes. Using the Silhouette score, the KPI interface.bond0.if_octets.rx presents the best clustering results (0.598 for 4 clusters) followed by the KPI interface.bond0.if_octets.tx (0.580 for 4 clusters). Using the Davies-Bouldin index, the KPI interface.bond0.if_octets.rx presents the best clustering results (0.34 for 12 clusters) followed by the KPI interface.bond0.if_octets.tx (0.37 for 13 clusters). Using the Calinski-Harabasz index, the KPI interface.bond0.if_octets.tx presents the best clustering results (726.3 for 4 clusters) followed by the KPI interface.bond0.if_octets.rx (687.9 for 4 clusters).

Consequently, we can conclude that the Network (interface) traffic KPIs (interface.bond0.if_octets.rx and interface.bond0.if_octets.tx) present the best clustering results over all 11 KPIs, providing 4 and 13 clustering, respectively. In order to decide which is the most adequate number of clusters for our dataset, i.e. the most adequate KPI, we have analyzed the time series decomposition of all the jobs per cluster. Figure 6 shows sample jobs from two different clusters A and B from the optimal result obtained with the KPI interface.bond0.if_octets.rx. Figure 7 also displayed the working nodes behaviors of each job. After our analysis, we concluded that this KPI (interface.bond0.if_octets.rx) is the one that shows a high percentage of jobs with similar trends and behavior.

The results of the second comparison conclude that, according to the Silhouette and Davies-Bouldin indexes, the best results are obtained applying hierarchical algorithms. However, and according to the Calinski-Harabasz index, K-means is the best option. Since we obtain the same conclusion in two out of three clustering validation methods, we consider that hierarchical algorithm is the most adequate for our purpose. Besides, Calinski-Harabasz index does not have an upper value level, so it is usually applied to compare different classifications with the same conditions, which reinforce our approach.

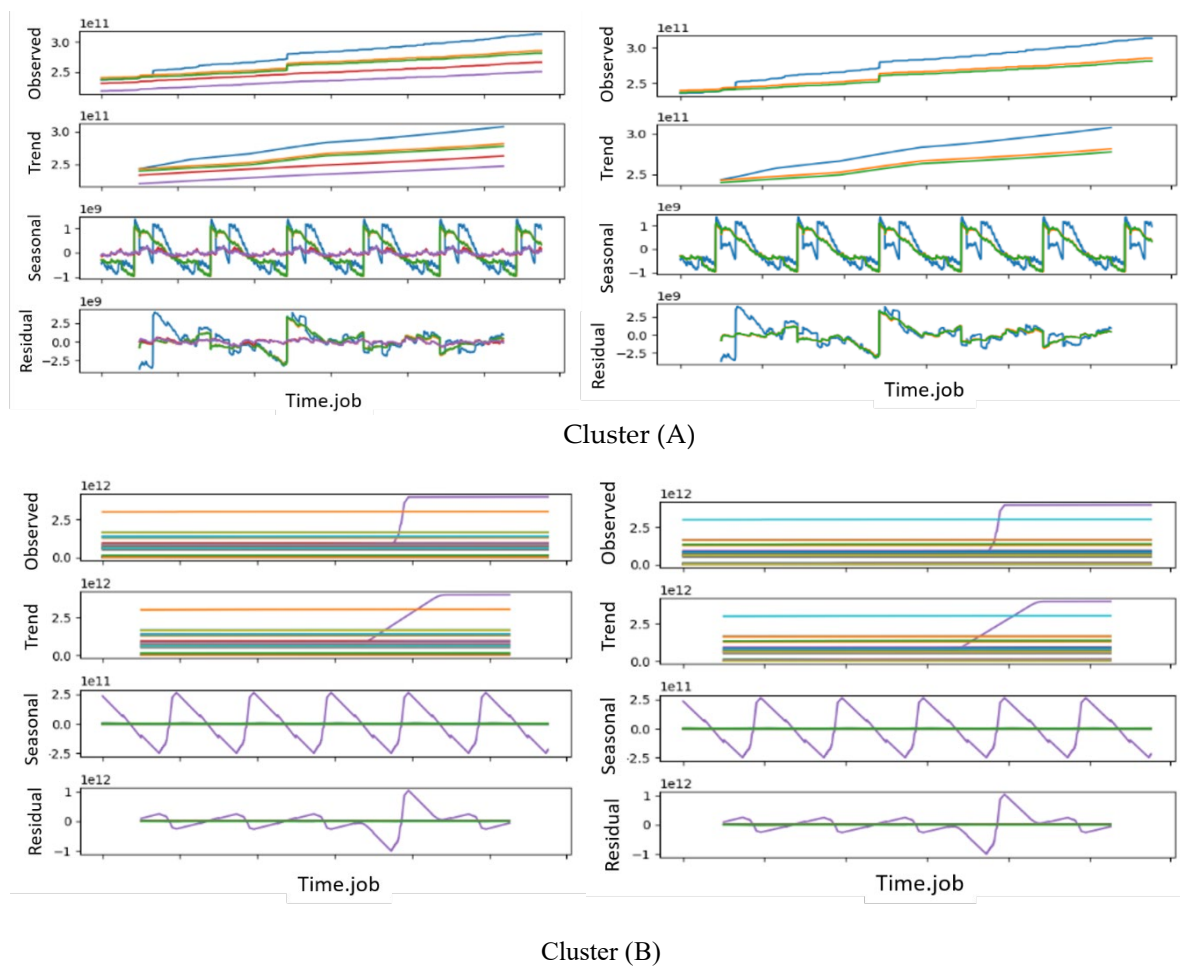

Cluster (A)

Cluster (B)

**Figure 6.** A time series decomposition of jobs from two different clusters of the KPI *interface.bond0.if_octets.rx*.

Finally, our results were validated by conducting a clustering experiment with a new dataset, which has confirmed that the Network (interface) traffic KPIs (interface.bond0.if_octets.rx and interface.bond0.if_octets.tx) show the best clustering results.

**7. Conclusions**

This study aimed to provide a methodology to cluster HPC jobs (non-operational) in order to automatically detect different types of jobs according to their performance. The job performance can be studied by using the KPI metrics provided by the HPC monitoring system. Our goal was also to select the most suitable or representative set of KPIs for clustering non-operational jobs according to their performance. Our analysis and validation were done by using a data set provided by the Supercomputing Center of Galicia (CESGA) that collected the information of the KPIs of 1,783 jobs from 1st June 2018 to 31st July 2018.

Considering the large amount of available KPIs (44,280), we have made a previous selection based on the advice from experts who work at CESGA. They provided us with 11 KPIs from the following categories: CPU usage, Memory usage, IPMI, System load and Network (interface) traffic.

We performed two different kinds of experiments in order to select the most suitable KPIs for clustering HPC jobs. The first experiment performed the clustering by combining the information gathered from the 11 KPIs, whereas the second one performed the individual clustering individually for each one of the 11 KPIs. Both experiments were done by using different clustering algorithms (K-means and agglomerative hierarchical algorithm), using different linkage methods (single-linkage, complete-linkage, average-linkage and Ward's method), and using different distance metrics (Euclidean, Manhattan and Cosine). In order to assess the quality of the obtained clusters, we have also used different indexes (Silhouette, Calinski-Harabasz and Davies-Boulding). Before performing the clustering, we have applied PCA in order to reduce the dimensionality of the data, without losing information, to reduce the computational load of the algorithms. Finally, a clustering experiment based only on the two selected KPIs (interface.bond0.if_octets.rx, and interface.bond0.if_octets.tx) was performed with the aim of validating our approach. For this, we have obtained a new dataset with 1,500 jobs (non-operational) from 1st August 2019 to 31st September 2019. The results confirmed our proposal.

Our analysis concluded that the clustering based on the joint information given by the 11 KPIs performed worse than the clustering based on the individual KPIs. What is more, the results showed that the information given by those KPIs belonging to the Network (interface) traffic, are the most adequate (interface.bond0.if_octets.rx and interface.bond0.if_octets.tx). The clusters obtained with the information of these KPIs showed the best quality in terms of cohesion and separation of HPC jobs. More specifically, the visualization of the KPI (interface.bond0.if_octets.rx) clusters showed a high percentage of jobs with similar trends. Therefore, our methodology can be applied to any data set with information about these two KPIs in order to obtain a good clustering and infer the number of types of non-operational jobs that run in the HPC system. The procedure is simple and offers a solution to some challenges faced in other experimentations [9,10,11] when dealing with similar unlabeled data with large dimensionality.

In our opinion, this clustering phase should be considered the first stage in a broader procedure to detect anomalies in HPC systems. In fact, we are currently working on analyzing this categorization. We consider that the obtained clusters would help to infer similar characteristics of the jobs belonging to each cluster that, definitively, could give information to detect those jobs whose performance is not the expected one and be able to early detect potential anomalies in the system. Finally, and although we have checked that the mechanism applied for dimensionality reduction (fixed PCA) supports a good percentage of retained information, we are working to improve this aspect. Since it was mentioned in the literature [25,27], the problem with the cost function used in PCA entails that there are retained large pairwise distances instead of focusing on retaining small pairwise distances, which is usually much more important. The solution given in [25] is defining a specific cost function based on a non-convex objective function. We are currently defining this new cost function using a larger dataset obtained from the same high performance computing center. We are also considering to use KPIs time series feature extraction in our clustering methodology. The extracted features statistical significance will be evaluated and analyzed by different state-of-art machine learning approaches to achieve our purpose.

**Author Contributions:** For research articles with several authors, a short paragraph specifying their individual contributions must be provided. The following statements should be used "Conceptualization, M.S.H, R.P.D.R and A.F.V.; methodology, M.S.H, R.P.D.R and A.F.V..; software, M.S.H., validation, M.S.H.; formal analysis, M.S.H, R.P.D.R and A.F.V.; investigation, M.S.H, R.P.D.R and A.F.V.; resources, R.P.D.R and A.F.V; data curation, M.S.H.; writing—original draft preparation, M.S.H, R.P.D.R and A.F.V.; writing—review and editing, M.S.H, R.P.D.R and A.F.V.; supervision, R.P.D.R and A.F.V; funding acquisition, R.P.D.R and A.F.V. All authors have read and agreed to the published version of the manuscript".

**Funding:** This work was supported by the European Regional Development Fund (ERDF) and the Galician Regional Government, under the agreement for funding the Atlantic Research Center for Information and Communication Technologies (AtlantTIC), and the Spanish Ministry of Economy and Competitiveness, under the National Science Program (TEC2017-84197-C4-2-R). The authors would also like to thank the Supercomputing Center of Galicia (CESGA) for their support and the resources for this research.

**Acknowledgments:** The authors would like to thank the European Regional Development Fund (ERDF) and the Galician Regional Government, under the agreement for funding the Atlantic Research Center for Information and Communication Technologies (AtlanTTIC), and the Spanish Ministry of Economy and Competitiveness, under the National Science Program (TEC2017-84197-C4-2-R). The authors would also like to thank the Supercomputing Center of Galicia (CESGA) for their support and the resources for this research.

**Conflicts of Interest:** The authors declare no conflict of interest

**Appendix A**

Table A1. Experiment two: results

| KPI | VM | K-means Euclidean | | Agglomerative Hierarchical | | | | | | | | | | | | | | | | |
|---|---|---|---|---|---|---|---|---|---|---|---|---|---|---|---|---|---|---|---|---|
| | | | | Euclidean | | | | | | | | Cosine | | | | | | Manhattan | | | |
| | | | | Average | | Complete | | Single | | Ward | | Average | | Complete | | Single | | Average | | Complete | | Single | |
| | | K | Score | K | Score | K | Score | K | Score | K | Score | K | Score | K | Score | K | Score | K | Score | K | Score | K | Score |
| 1 | C | 6 | 70.13 | 5 | 53.53 | 12 | 28.02 | 16 | 11.59 | 11 | 13.89 | 10 | 32.07 | 8 | 51.30 | 15 | 15.84 | 8 | 37.34 | 14 | 41.21 | 7 | 4.98 |
| | D | 8 | 1.59 | 5 | 1.940 | 11 | 1.99 | 5 | 1.34 | 10 | 1.45 | 9 | 0.78 | 5 | 1.11 | 15 | 0.62 | 7 | 0.77 | 7 | 1.46 | 8 | 0.62 |
| | S | 9 | 0.251 | 7 | 0.137 | 5 | 0.14 | 16 | 0.034 | 9 | 0.24 | 10 | 0.24 | 8 | 0.20 | 15 | 0.12 | 7 | 0.29 | 7 | 0.19 | 8 | 0.18 |
| 2 | C | 7 | 51.78 | 11 | 22.05 | 20 | 22.73 | 11 | 10.02 | 6 | 23.68 | 15 | 21.85 | 5 | 52.12 | 6 | 29.01 | 5 | 36.48 | 7 | 34.92 | 22 | 9.65 |
| | D | 4 | 1.40 | 17 | 1.57 | 11 | 2.08 | 10 | 1.32 | 3 | 1.37 | 11 | 0.61 | 10 | 1.08 | 6 | 0.49 | 5 | 0.78 | 10 | 1.34 | 7 | 0.502 |
| | S | 4 | 0.212 | 11 | 0.12 | 11 | 0.10 | 11 | -0.07 | 8 | 0.21 | 15 | 0.23 | 5 | 0.21 | 6 | 0.38 | 5 | 0.36 | 8 | 0.18 | 22 | 0.02 |
| 3 | C | 6 | 69.48 | 6 | 22.85 | 5 | 25.75 | 6 | 22.56 | 4 | 33.21 | 13 | 35.21 | 3 | 90.84 | 7 | 33.48 | 5 | 66.19 | 14 | 32.55 | 4 | 57.69 |
| | D | 5 | 1.347 | 11 | 1.94 | 8 | 2.28 | 6 | 1.50 | 5 | 1.10 | 9 | 0.99 | 20 | 1.09 | 6 | 0.55 | 7 | 0.97 | 5 | 1.31 | 6 | 0.75 |
| | S | 5 | 0.254 | 13 | 0.11 | 5 | 0.09 | 12 | 0.03 | 5 | 0.25 | 13 | 0.25 | 6 | 0.25 | 7 | 0.25 | 9 | 0.20 | 5 | 0.19 | 6 | 0.08 |
| 4 | C | 4 | 687.9 | 9 | 92.54 | 9 | 111.2 | 9 | 84.32 | 6 | 110.3 | 13 | 341.4 | 6 | 90.5 | 11 | 184.7 | 4 | 615.8 | 8 | 366.6 | 17 | 151.4 |
| | D | 8 | 0.96 | 4 | 1.02 | 12 | 1.92 | 6 | 0.85 | 8 | 0.92 | 9 | 0.74 | 7 | 0.80 | 12 | 0.34 | 9 | 0.46 | 30 | 0.83 | 22 | 0.37 |
| | S | 8 | 0.39 | 12 | -0.02 | 9 | 0.024 | 9 | 0.01 | 8 | 0.40 | 9 | 0.53 | 7 | 0.48 | 11 | 0.45 | 4 | 0.59 | 6 | 0.53 | 17 | 0.39 |
| 5 | C | 4 | 726.3 | 18 | 58.29 | 14 | 86.9 | 6 | 66.5 | 6 | 77.8 | 11 | 366.4 | 4 | 638.6 | 6 | 300.5 | 5 | 514.1 | 6 | 600.2 | 3 | 440.5 |
| | D | 8 | 1.04 | 19 | 1.00 | 29 | 1.22 | 18 | 0.83 | 6 | 1.09 | 6 | 0.56 | 4 | 0.603 | 13 | 0.37 | 6 | 0.55 | 7 | 0.96 | 5 | 0.53 |
| | S | 8 | 0.37 | 10 | 0.007 | 6 | 0.01 | 9 | 0.01 | 6 | 0.351 | 4 | 0.70 | 4 | 0.56 | 6 | 0.39 | 4 | 0.58 | 7 | 0.387 | 5 | 0.56 |
| 6 | C | 11 | 62.27 | 6 | 34.98 | 15 | 21.92 | 3 | 45.88 | 5 | 30.42 | 6 | 54.48 | 9 | 62.28 | 12 | 25.59 | 8 | 58.50 | 9 | 59.09 | 4 | 59.33 |
| | D | 4 | 1.09 | 6 | 1.88 | 9 | 2.15 | 4 | 1.13 | 4 | 1.10 | 13 | 0.74 | 5 | 1.00 | 6 | 0.55 | 4 | 0.60 | 4 | 0.89 | 4 | 0.49 |
| | S | 4 | 0.28 | 6 | 0.07 | 7 | 0.35 | 5 | 0.21 | 4 | 0.28 | 6 | 0.31 | 5 | 0.30 | 5 | 0.31 | 7 | 0.32 | 9 | 0.24 | 4 | 0.32 |
| 7 | C | 10 | 59.27 | 6 | 38.27 | 7 | 22.71 | 4 | 45.52 | 5 | 50.4 | 7 | 46.62 | 11 | 51.31 | 5 | 45.19 | 6 | 60.70 | 10 | 53.21 | 6 | 38.22 |
| | D | 4 | 1.63 | 9 | 1.57 | 8 | 1.97 | 6 | 1.42 | 4 | 1.03 | 19 | 0.70 | 6 | 0.94 | 6 | 0.51 | 9 | 0.72 | 5 | 0.87 | 9 | 0.54 |
| | S | 4 | 0.31 | 7 | 0.06 | 5 | 0.06 | 5 | -0.03 | 4 | 0.296 | 7 | 0.33 | 11 | 0.25 | 5 | 0.31 | 9 | 0.32 | 7 | 0.26 | 6 | 0.30 |
| 8 | C | 13 | 54.26 | 4 | 61.43 | 16 | 28.88 | 8 | 27.04 | 11 | 60.86 | 7 | 46.49 | 5 | 76.66 | 4 | 59.1 | 10 | 47.57 | 10 | 52.39 | 9 | 24.82 |
| | D | 9 | 1.51 | 7 | 1.66 | 7 | 2.31 | 6 | 1.17 | 4 | 1.05 | 17 | 0.74 | 12 | 1.02 | 4 | 0.46 | 18 | 0.74 | 11 | 1.05 | 9 | 0.54 |
| | S | 4 | 0.29 | 9 | 0.11 | 7 | 0.08 | 4 | 0.008 | 4 | 0.29 | 7 | 0.32 | 11 | 0.25 | 4 | 0.37 | 9 | 0.294 | 10 | 0.26 | 9 | 0.23 |
|    | C | 10 | 60.65 | 5  | 43.82 | 7  | 49.37 | 14 | 16.54  | 11 | 56.54 | 11 | 40.54 | 5  | 82.12 | 5  | 34.38 | 11 | 38.15 | 6  | 78.59 | 7  | 24.53 |
| 9  | D | 13 | 1.39  | 7  | 1.21  | 7  | 2.12  | 4  | 1.29   | 6  | 1.09  | 6  | 0.684 | 5  | 0.99  | 5  | 0.49  | 13 | 0.76  | 4  | 0.97  | 7  | 0.50  |
|    | S | 4  | 0.34  | 5  | 0.05  | 9  | 0.17  | 14 | 0.009  | 6  | 0.319 | 11 | 0.32  | 5  | 0.34  | 5  | 0.34  | 11 | 0.32  | 5  | 0.34  | 7  | 0.28  |
|    | C | 5  | 73.93 | 5  | 45.75 | 4  | 50.93 | 8  | 17.27  | 10 | 56.25 | 6  | 38.23 | 6  | 62.1  | 17 | 14.04 | 9  | 35.37 | 13 | 39.82 | 6  | 4.25  |
| 10 | D | 5  | 1.71  | 6  | 2.03  | 5  | 1.96  | 9  | 1.40   | 5  | 1.71  | 5  | 0.74  | 22 | 1.21  | 5  | 0.59  | 6  | 0.74  | 5  | 1.27  | 6  | 0.58  |
|    | S | 6  | 0.19  | 5  | 0.088 | 5  | 0.13  | 5  | -0.04  | 9  | 0.23  | 6  | 0.26  | 7  | 0.16  | 7  | 0.19  | 14 | 0.23  | 5  | 0.20  | 4  | 0.21  |
|    | C | 10 | 37.98 | 5  | 36.32 | 10 | 17.75 | 8  | 11.60  | 10 | 40.35 | 10 | 21.83 | 8  | 35.19 | 5  | 38.38 | 8  | 25.13 | 9  | 31.11 | 18 | 11.71 |
| 11 | D | 9  | 1.83  | 6  | 1.99  | 8  | 2.19  | 15 | 1.15   | 5  | 1.27  | 4  | 0.57  | 5  | 1.21  | 7  | 0.56  | 4  | 0.57  | 8  | 1.72  | 4  | 0.55  |
|    | S | 5  | 0.16  | 5  | 0.13  | 4  | 0.07  | 13 | -0.03  | 4  | 1.18  | 7  | 0.28  | 5  | 0.30  | 5  | 0.31  | 6  | 0.26  | 4  | 0.13  | 9  | 0.12  |